\documentclass[aps,prl,superscriptaddress,twocolumn]{revtex4}
\usepackage{amsmath}
\usepackage{graphicx}
\usepackage{color}

\begin{document}

\newcommand{\beq}{\begin{equation}}
\newcommand{\eeq}{\end{equation}}
\newcommand{\beqa}{\begin{eqnarray}}
\newcommand{\eeqa}{\end{eqnarray}}
\newcommand{\ben}{\begin{enumerate}}
\newcommand{\een}{\end{enumerate}}
\newcommand{\hs}{\hspace{0.5cm}}
\newcommand{\vs}{\vspace{0.5cm}}

\title{Detecting Classical Phase Transitions with Renyi Mutual Information }

\author{Jason Iaconis}
\affiliation{Department of Physics and Astronomy, University of Waterloo, Ontario, N2L 3G1, Canada}
\affiliation{Department of Physics, University of California, Santa Barbara, CA, 93106-9530}

\author{Stephen Inglis}
\affiliation{Department of Physics and Astronomy, University of Waterloo, Ontario, N2L 3G1, Canada}

\author{Ann B. Kallin}
\affiliation{Department of Physics and Astronomy, University of Waterloo, Ontario, N2L 3G1, Canada}

\author{Roger G. Melko}
\affiliation{Department of Physics and Astronomy, University of Waterloo, Ontario, N2L 3G1, Canada}
\affiliation{Perimeter Institute for Theoretical Physics, Waterloo, Ontario N2L 2Y5, Canada}

\date{\today}

\begin{abstract}
By developing a method to represent the Renyi entropies via a replica-trick 
on classical statistical mechanical systems,
we introduce a procedure to calculate the Renyi Mutual Information (RMI) in any Monte Carlo simulation.
Through simulations on several classical models, we demonstrate that the RMI 
can detect finite-temperature critical points, and even identify their universality class, 
without knowledge of an order parameter or other thermodynamic estimators.  
Remarkably, in addition to critical points mediated by symmetry breaking, the RMI is
able to detect topological vortex-unbinding transitions, as we explicitly demonstrate on 
simulations of the  XY model.
\end{abstract}

\maketitle

{\em Introduction --}
The universality and importance 
of the concept of information is exploited widely in mathematics and the physical sciences.
Information sets the fundamental limits in communication (or uncertainty), regardless of system, technology, or physical material.
Shannon was the first to quantify information using the concept of {\it entropy} -- a quantity that has its roots in thermodynamics \cite{Shannon_1948}.  It is therefore not
surprising that deep ties exist between the measurement of thermodynamic quantities and concepts associated with information theory.  

There is already a rich cross-fertilization between ideas in condensed matter physics and the information sciences.  Most recently, information measures have been used to quantify ``hidden'' correlations in materials -- exchanges of information that can occur between two parts of a system that are {\em not} manifest in traditional condensed-matter estimators (such as correlation functions) \cite{PhysRevLett.100.070502}.  For example, in spin liquid phases, correlation functions can rapidly decay as a function of spatial separation; however, due to constraints, hidden correlations exist across vast distances of the sample \cite{topo_ord}.  These can be manifest in entropy quantities measuring the amount of communication between two regions of the sample -- resulting in a practical estimator for, among other things, topological order in condensed matter systems \cite{LW,KP}.

Phase transitions offer another testing ground for the use of information quantities in condensed matter systems.  Critical points are associated with a diverging correlation length, suggesting the existence of long-range channels for information transfer.  
However, it is not obvious that measurable quantities associated with this information can be exploited to tell us anything about these phase transitions.

In this paper, we examine the Renyi Mutual Information (RMI), 
a measure that quantifies the amount of information contained in some region of a statistical mechanical system, about the rest of the system.
Numerical measurements of classical mutual information \cite{Wilms1,Wilms2} typically calculate the 
reduced density matrix explicitly, a computationally expensive task, and use that to calculate the von Neumann entropy directly.
The RMI on the other hand is easily measured in standard  Monte Carlo routines
via a {\it replica trick} \cite{replicaMC1,Buividovich}, rather than calculation of a reduced density matrix, making it immediately amenable to measurement on a vast number of models of interest to condensed matter, biophysics, and physical chemistry.  
We show that this RMI can be used in a practical way to identify phase transitions through finite-size scaling analysis on lattices of different sizes, {\it without} knowledge of an order parameter or any other thermodynamic quantity.  We use the standard two-dimensional Ising model as a test case, demonstrating universality when results are compared to vastly different (even quantum mechanical \cite{Roger_MI}) models that exhibit the same universality class.  Finally, we establish the ability of the RMI to detect the Berezinskii-Kosterlitz-Thouless (BKT) transition, without relying on knowledge of any thermodynamic estimator such as the spin stiffness.  This suggests the power of the RMI for detecting hidden transitions in a variety of other statistical mechanical models in the future.

{\em Information, entropy, and the replica trick --}
Given a random variable $X$ one can quantify its associated uncertainty, or equivalently, the amount of information one is missing by not knowing the state of $X$.
There are various ways that this information can be embodied,
for example in the generalized Renyi entropies \cite{renyi},
\beq \label{renyi}
S_{\alpha}(X) = \frac{1}{1-\alpha} \ln \left( \sum_{i \in X}  p_i^{\alpha} \right),
\eeq
where $p_i$ is the probability of outcome $X=i$.
Taking the limit $\alpha \rightarrow 1$, one recovers Shannon's familiar entropy, $S_1 = -\sum_i p_i \ln(p_i)$, which
can be related to the thermodynamic entropy of a statistical mechanical system $S = \ln \Omega$, where $\Omega$ is the number of microstates, assuming all occur with equal probability.
This relationship is often exploited in the study of real physical systems through microscopic statistical mechanical models.

In such systems, one may also ask how much knowledge of a subsystem (call it $A$) is possible, assuming complete knowledge of 
another subsystem $B$.  Correlation functions are a common example
that incompletely quantify this knowledge.  In this paper, let us instead define a spatial subregion $A$ as the complement of $B$, so that
$A \cup B$ is the complete system;
we could say that $p_{i_A}$ is the probability of state ${i_A}$ occurring in region $A$.  
The state $i_A$ can be any classical (or even quantum) state: we will restrict our focus mainly to classical spin systems.

From Boltzmann, the probability of a state occurring is 
$p_{i_A,i_B} = e^{-\beta E(i_A,i_B)}/{Z}$,
where $E(i_A,i_B)$ is the energy associated with states $i_A$ in region $A$ and $i_B$ in region $B$, and $Z= \sum_{i_A,i_B} e^{-\beta E(i_A,i_B)}$ is the partition function.  
To obtain only the probability of a certain state in $A$, we can instead sum over all possible states in $B$,
$
p_{i_A} = \sum_{i_B}e^{-\beta E(i_A,i_B)}/{Z} .
$
The appropriate sums of $p_{i_A,i_B}$ or $p_{i_A}$, raised to the power $\alpha$, give the Renyi entropies in a straightforward way.
In equilibrium statistical mechanical systems,
estimators based on sums such as these are commonly calculated using importance sampling techniques, i.e. Markov chain Monte Carlo.

In this paper we mostly restrict our discussion to the {\it second} Renyi entropy,
which requires sums of the probabilities $p_{i_A,i_B}$ or $p_{i_A}$ squared.  
For example, to get the entropy of region $A$ we use,
$p_{i_A}^2 = \left(\sum_{i_B}e^{-\beta E(i_A,i_B)}\right)\left(\sum_{j_B}e^{-\beta E(i_A,j_B)}\right)/Z^2$,
leading to 
\begin{eqnarray}
S_2(A) &=& -\ln \left(Z^{-2}\sum_{i_A} \sum_{i_B} \sum_{j_B} e^{-\beta \left( E(i_A,i_B) + E(i_A,j_B) \right)} \right),  \nonumber \\
&=& - \ln (Z[A,2,T]) + 2 \ln(Z[T]), \label{S_2A}
\end{eqnarray}
where we have defined a ``replicated" partition function $Z[A,2,T]$\cite{roger_multisheet_sse}, which can be sampled via a Monte Carlo simulation procedure described in the next section.

Note that, while the Renyi entropies tell us about uncertainty in the full system or part of the system, they do not reveal the correlations or information {\it between} two regions of a system.
For this we introduce the RMI,
\beq \label{MI}
I_{\alpha}(A;\! B) = S_{\alpha}(A) + S_{\alpha}(B) - S_{\alpha}(A \cup B).
\eeq
This measure defines in a precise way the information that a full knowledge of $B$ gives us about $A$, or vice versa.
From \eqref{S_2A}, the RMI can be related to a difference in free energies.  
Note, the free energy in a typical condensed-matter system with a $D-1$ dimensional boundary generally \cite{onsager_transfer,Izmailian,Xintian} behaves as
$F = a L^D f(T) + b \ell f_B(T)$, where $f$ and $f_B$ are the size-independent free energy densities, and $\ell \propto L^{D-1}$.
The thermodynamic behavior of the RMI, which is constructed to cancel contributions arising from the bulk, is determined by this boundary free energy -- restricting it to at most ``area law" scaling $I_{\alpha} \propto \ell$ \footnote{This was discussed previously for the classical $I_1$ in Ref.~\cite{PhysRevLett.100.070502}, 
based on the fact that correlations in classical thermal states are localized at the boundary.}.
As we will see next, subleading corrections to this area law make the RMI an extremely useful tool for detecting phase transitions in finite-size systems.

\begin{figure}
\includegraphics[width=0.9\columnwidth]{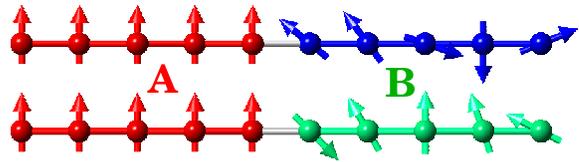}
\caption{ (color online).  A representation of the replicated partition function, $Z[A,2,T]$, used for calculating $S_2(A)$ in a one-dimensional system with 10 spins.  In region $A$ (the left 5 spins), replicas are constrained to always be identical.  In region $B$, configurations are sampled independently in the two replicas.
The constraints on spins in $A$ effectively halve its temperature, such that when the replicated system has temperature $T_c < T < 2T_c$, spins in this region are below-critical, while in $B$ the spins are above-critical (illustrated).
\label{S2}}
\end{figure}

{\em Thermodynamic behavior --}
A key observation that enables the widespread utility of the RMI for integer $\alpha \ge 2$ is that,
for any physical system, the calculation of $S_{\alpha}(A)$
can be accomplished via the replicated system geometry, $Z[A,\alpha,T]$ (Fig.~\ref{S2}).
For $S_2(A)$, the form of the Boltzmann factor effectively constrains states in $A$ to be equal between the two replicas, while states in $B$ are unconstrained between replicas.
This {\it replica trick} \cite{roger_multisheet_sse} leads to the practical method for measurement of $S_2$ in Monte Carlo simulations discussed in the next section.  In addition, it facilitates the general understanding of the RMI in the thermodynamic limit, as we now discuss.

We observe that, in addition to the general expectation that the RMI contains an ``area law'' term, 
important subleading constant corrections may occur.
In the simplest case, where $\Omega_0$ symmetry-broken ground states exist at a temperature far below $T_c$,
both $Z[T]=\Omega_0$ and $Z[A,2,T] =\Omega_0$.  Generalizing Eq.~\eqref{S_2A}, 
the RMI is then,
\begin{eqnarray}
I_{\alpha} (A;\! B) &=& \frac{1}{1-\alpha} \Bigl( \ln (Z[A,\alpha,T]) + \ln (Z[B,\alpha,T])  \Bigr.  \label{below_tc}  \\
 &-& \Bigl. \alpha \ln(Z[T]) - \ln (Z[A \cup B,\alpha,T]) \Bigr)  
= \ln\Omega_0, \nonumber
\end{eqnarray}
i.e. some positive constant independent of Renyi index.
When $T>0$ and fluctuations are included, this positive constant is added to the area law term discussed above.

Next, in the intermediate temperature range $T_c < T < \alpha T_c$,
the ``unconnected'' region ($B$) of the replicated system is above criticality, 
whereas the ``connected'' part ($A$) of the simulation is effectively below $T_c$ due to the constraint on $A$.
If each above-critical degree of freedom in the system can realize $\sigma$ different states, 
course-graining on the length scale of the correlation length $\xi$ results in a partition function of 
$Z[T] =  \sigma^{N/ \xi^D}$.
The replicated partition function, on the other hand, has a reduction in the number of accessible states
since the $\alpha$ above-critical regions are connected to a below-critical region through the boundary of length $\ell$, which eliminates $\mathcal{O}(\ell \xi)$ lattice degrees of freedom in each of the unconstrained regions, giving
$Z[A,\alpha,T] =  \Omega_0 \sigma^{(\alpha N_B - \alpha \ell \xi)/ \xi^D}$. 
Eq.~\eqref{below_tc} then gives
\begin{align}
I_{\alpha} (A;\! B) = & \frac{1}{1-\alpha} \Bigl( \ln\Omega_0\sigma^{(\alpha N_B - \alpha \ell \xi)/\xi^D} +\nonumber  \\
& \ln\Omega_0\sigma^{(\alpha N_A - \alpha \ell \xi)/\xi^D} - \alpha \ln \sigma^{N/\xi^D} 
- \ln \Omega_0 \Bigr)\nonumber \\
= & \frac{1}{1-\alpha} \Bigl( \ln\Omega_0 - 2 \alpha \ell/\xi^{D-1} \ln\sigma \Bigr) . \label{betweenTc}
\end{align}
This gives a constant part of the RMI, $c = -\ln\Omega_0/(\alpha-1)$ and a positive area law part $\ell (2 \alpha \ln\sigma/(\alpha - 1)\xi^{D-1})$.

Importantly, we see that the constant part of the RMI changes sign as we pass through $T_c$.
It can be seen in our finite-size Monte Carlo data that the presence of this constant $c$ will cause the $I_{\alpha}/\ell$ curves to ``fan out'' away from $T_c$ for different $\ell$, while $c=0$ precisely at $T_c$, produceing a striking crossing in the curves.

{\em Monte Carlo Algorithm --} 
The form of the replica trick suggests a straightforward way to measure the Renyi entropy using a modified simulation geometry (Fig.~\ref{S2}).
The constraint on region $A$ demands that, to be accepted, an update must affect states on the same physical lattice site in all replicas -- effectively reducing the temperature of region $A$ by a factor of $\alpha$, as discussed above.

Using this modified simulation we can generate states according to the probability $p_{i_A}^2$.
In general, generating states from a partition function via Monte Carlo does not allow direct calculation of the partition function (or the free energy) itself in an efficient manner.
To overcome this, one approach used previously~\cite{roger_multisheet_sse} is to integrate the energy estimator starting from $T \rightarrow \infty$, obtaining the Renyi entropy at some finite $T$ from both a replicated simulation and an unreplicated simulation.
This necessitates a schedule of simulations over a large range of temperatures to gather enough detailed knowledge of the energy for an accurate integration
\footnote{
Alternatively, it may be possible to use methods able to directly extract the ratio of partition functions at a given temperature, as done
previously for the analogous quantum method~\cite{roscilde_qmc} }.

{\em Results on models -- }
Using conventional Monte Carlo simulations of several models, we demonstrate the use of the second RMI to detect finite-temperature phase transitions.
In the following, we use $I_2(A;\! B)$ where $A$ and $B$ are complementary regions, each defined as an $L \times L/2$ cylinder embedded in the $L \times L$ torus.

\begin{figure}
\includegraphics[width=\columnwidth]{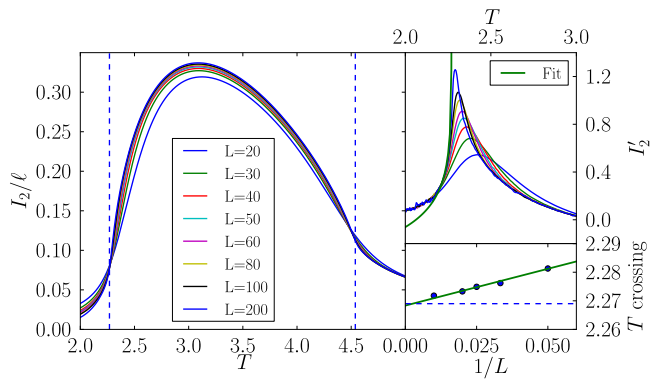}
\caption{(color online). Left: The RMI per boundary length ($I_2/\ell$) as a function of temperature for the Ising model. Dashed lines indicate $T_c$ and $2T_c$.
Upper right: the first derivative of the RMI for each system size as a function of temperature, with a fit to a logarithmic divergence shown in Eq.~\eqref{ising_log}.
Lower right: the temperature of the lower crossings of the RMI for sizes $L$ and $2L$ as a function of $1/L$. Dashed line indicates $T_c$. Finite size scaling gives us $T_c = 2.2683(17)$.
\label{ising}}
\end{figure}

The first model we examine is the classical Ising model on a two-dimensional square lattice
$
H = -J \sum_{\langle i j \rangle} S^z_i S^z_j,
\label{ising_eq}
$ 
where $S^z_i = \pm {1}/{2}$.
There  exists a transition to an ordered phase at a temperature $T_c/J = 2/\log(1+\sqrt{2}) \approx 2.269$~\cite{onsager_transfer}, which
offers us the simplest test for the RMI.
Figure~\ref{ising} illustrates $I_2(A;\! B)/ \ell$ where
$\ell$ is the length of the boundary between the regions (in all cases hereafter, $\ell = 2L$).
Close inspection indicates approximate crossings of the $I_2(A;\! B)/ \ell$ curves at $T_c$ and $2T_c$.
Examining the crossings as we move to larger system sizes we see that they extrapolate towards the transition temperature. 
Knowing that the correlation length at the Ising transition behaves as $\xi \sim |t|^{-\nu}$ with $t=\lvert T-T_c\rvert / T_c$ and $\nu=1$, one can derive a  finite-size scaling behavior for the crossing temperature $T(L) - T_c \propto 1/L$.  This is confirmed by the data in the lower panel of Fig.~\ref{ising}.

Remarkably, this behavior mimics the crossing seen previously in the analogous RMI quantity based on the 
{\it entanglement} entropies in a quantum spin-1/2 XXZ model~\cite{stephen_wl,roger_multisheet_sse}.  That model also realizes a
finite-temperature critical point, in the 2D Ising universality class (however at a different $T_c$, which is non-universal).  
There, it has been argued that the crossings were a manifestation of criticality, with the scaling form \cite{Roger_MI},
\beq 
I_2(A;\! B) = [ c_1(t) + \phantom \cdot t \log t ] \cdot \ell + c_2(t)  + \mathcal{O}(1/\ell) \label{ising_log}.
\eeq
It is important to note that $c_1(t)$ can be polynomial in $t$, where the strict ``area law'' at $t=0$ is caused entirely by its constant piece
crossing zero ($\Omega_0=2$ in Eqs.~(\ref{below_tc}) and (\ref{betweenTc})). 
Divergences in derivatives of $I_2(A;\! B)$ are caused by the $t \log t$ contribution.
This term is known from the $t \log t$ divergence in the boundary free energy as $t \to 0$ for the square lattice Ising model with a field applied to the boundary on an infinite half plane \cite{mccoy_wu,me_fisher};
we can thus use it as a test of universality in this model.
In Fig.~\ref{ising}, $\partial I_2 /\partial t$ shows the predicted $\log t $ singularity, confirming that this critical point lives within the the 2D Ising universality class.

\begin{figure}[t]
\includegraphics[width=\columnwidth]{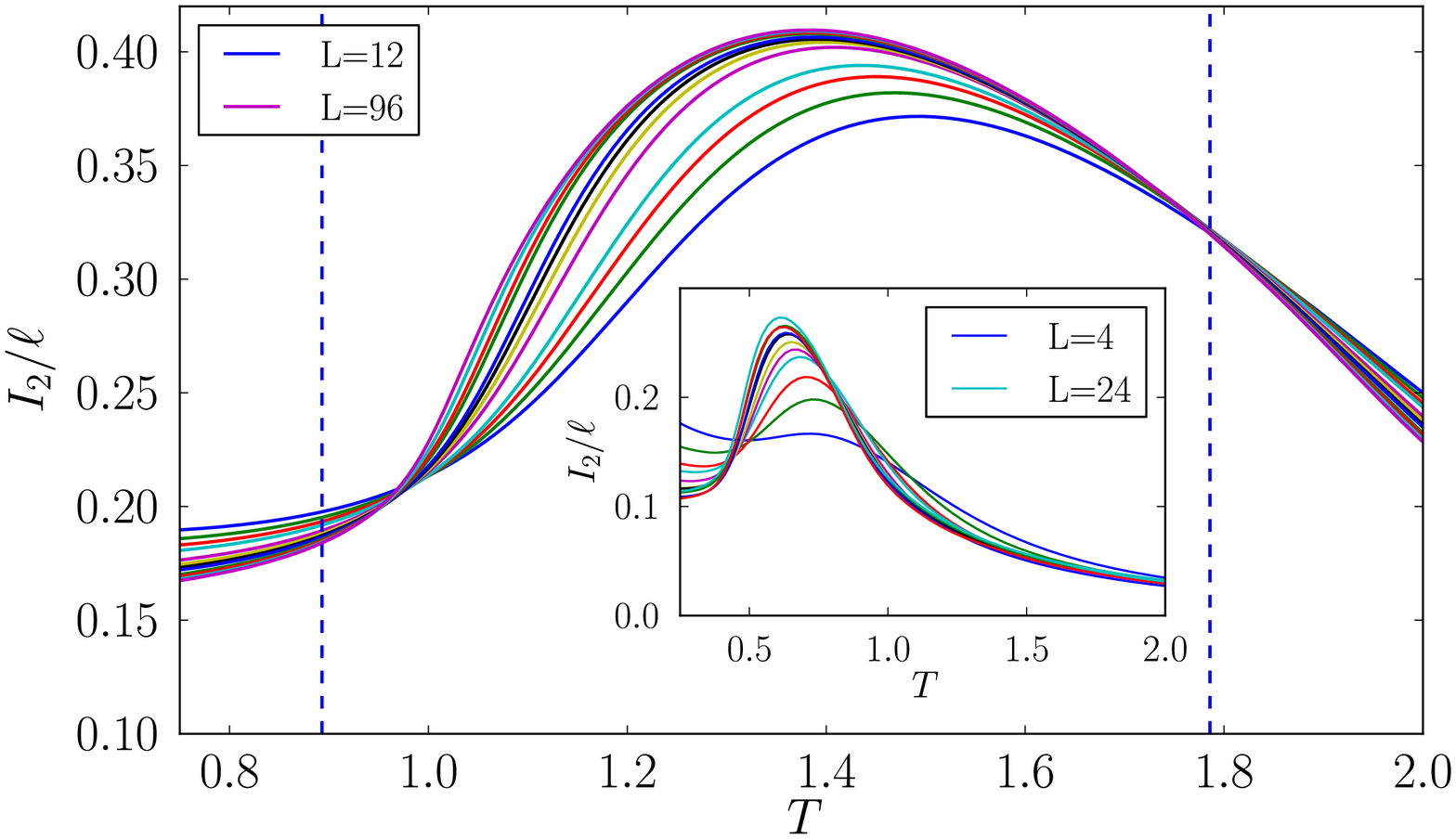}
\caption{(color online). The RMI per boundary length ($I_2/\ell$) as a function of temperature for the classical XY model, dashed lines indicate $T_{BKT}$ and $2T_{BKT}$.
Inset: Data for the quantum spin-1/2 XY model, obtained using a Wang-Landau technique~\cite{stephen_wl}.
Note that classically we reach much larger systems (only smallest and largest sizes are labeled).
\label{xy_cross}}
\end{figure}

We turn now to Monte Carlo simulations of the classical 2D XY model on a square lattice,
$
H = -J_{\rm XY} \sum_{\langle i j \rangle} \cos \left({ \theta_i - \theta_j }\right) . \label{xy_ham}
$
This is a model with continuous spin variable $0 \le \theta_i < 2 \pi$ that undergoes a BKT~\cite{KT_1} transition from a phase with free vortices to one with bound vortex-antivortex pairs.
Numerically, the detection of the BKT transition 
is much more subtle than the Ising phase transition,
since it has no local order parameter 
and standard scaling theory on thermodynamic estimators (such as the specific heat) does not work.
In 2D, this is circumvented by measuring the  
``spin stiffness'' and making use of a special universal jump condition, $T_{\rm BKT} = \pi \rho_s/2$, \cite{NK_jump} -- a procedure that has found $T_{BKT} = 0.89294(8)$ \cite{hasenbusch_TKT}. 
However, one may wonder if the RMI 
can detect this  phase transition without any need for such specialized measurements.

To address this, we examine the RMI for the classical XY model, shown in Figure~\ref{xy_cross}.
One can clearly see the development of crossings in the quantity $I_2/\ell$ near the value of $T_{\rm BKT}$ and 
$2 T_{\rm BKT}$ -- 
strong indication that universal scaling is coming into play.  
To examine this further, in the inset
we illustrate the RMI for a completely different model, the quantum
XY model 
$H = -t \sum_{\langle i j \rangle}  \left({ b^{\dagger}_i b^{\phantom\dagger}_j + b^{\phantom\dagger}_i b^{\dagger}_j }\right)
$, computed using Stochastic Series Expansion QMC
where the quantum RMI is generated using a broad histogram approach~\cite{stephen_wl}.
In this case, similar crossings appear at the (non-universal) $T_{\rm BKT} \approx 0.343$, giving strong
evidence in support of the universality of our result that the RMI can detect the BKT transition.

\begin{figure}
\includegraphics[width=\columnwidth]{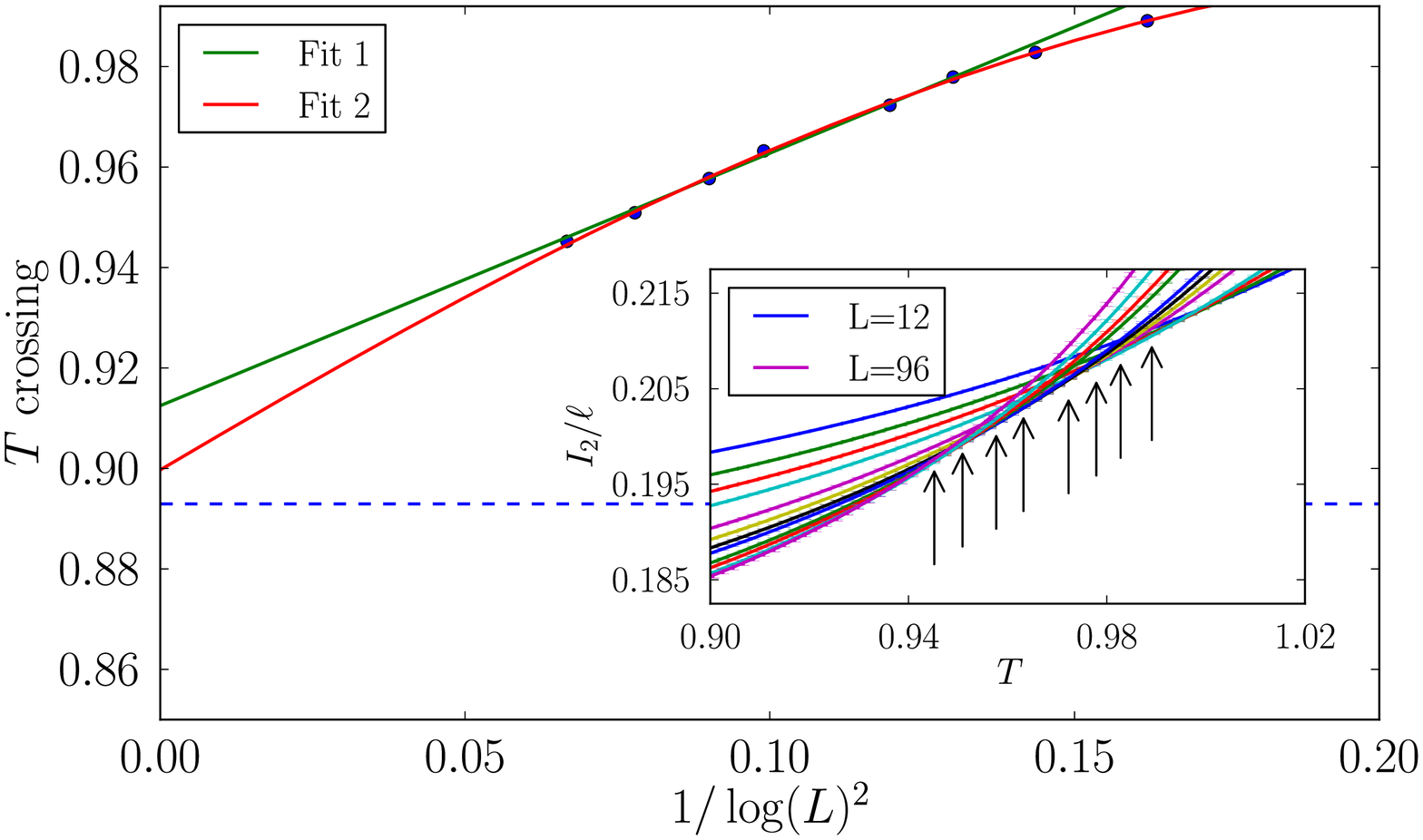}
\caption{(color online). The crossing of the RMI for the classical XY model between sizes $L$ and $2L$ as a function of $1/\log(L)^2$.
Fit 1 fits the points to Eq.~\eqref{xy_1order} assuming $c=0$ and using the largest six systems while Fit 2 uses the full equation and all of the data.
They give $T_{BKT}$ estimate of $0.912(4)$ and $0.899(9)$, respectively.
Inset: close-up of the lower crossing (only smallest and largest sizes are labeled). The arrows indicate the eight crossing values used in the main figure.
\label{xy_scale}}
\end{figure}

However, unlike the relatively ``clean'' crossing of the 2D Ising critical point, these
crossings have a larger finite-size scaling component.
 Figure~\ref{xy_scale} contains a detailed finite-size scaling analysis for the XY model.
This shows fits to two finite-size scaling forms which are derived by setting the correlation length near the critical temperature, $\xi = e^{-b / \sqrt{t}} (1 + \mathcal{O}(t) ) $  \cite{0305-4470-13-2-024}, equal to the linear system size $L / L_0$ with $t = (T - T_{BKT})/T_{BKT}$.  Then by relating the temperature $T$ with the finite-size transition temperature $T(L)$, we derive: 
\beqa
 T_{BKT}(L) = T_{BKT}(\infty) \left [ 1 + \frac{b^2}{\log^2(L/L_0)} + \frac{c}{\log^5(L/L_0)} \right ]   \label{xy_1order} .
\eeqa
Note, typical finite-size scaling analyses set the coefficient $c$ equal to zero, effectively ignoring subleading $\mathcal{O}(t)$ corrections to the correlation length.  Using this finite-size scaling form, the data for the crossing of $I_2/\ell$ convincingly approaches $T_{\rm BKT}$ in the limit of $L \rightarrow \infty$. 

{\em Discussion --}
The Renyi Mutual Information (RMI) is a quantity able to detect all correlations in a physical system, even those
missed by traditional connected correlation functions.
We have introduced a practical method to calculate the RMI using a modification of standard Monte Carlo techniques for classical statistical mechanical systems. 
We demonstrated that the RMI associated with the second Renyi entropy, $I_2$, is able to identify both conventional critical points, as well as the BKT transition where standard scaling theory breaks down.  A straight-forward finite-size scaling analysis of $I_2$ is sufficient to identify each phase transition, without knowledge of an order parameter, broken symmetry, or critical theory.

The ease of implementation of the RMI measurement in any standard Monte Carlo routine could stimulate adoption to simulation studies in many fields of the physical sciences and beyond.  
The RMI will likely find great utility in many classes of classical models, 
such as generalized XY models with ``hidden'' transitions \cite{Paul}, or
loop \cite{3Dloop,ScottLoop} and dimer models \cite{Dimer1,Dimer2} where universal properties of exotic criticality may be manifested \cite{Brian}.  
Also particularly pressing is the question \cite{ClaudioGlass} of whether RMI can detect unconventional transitions in disordered or glassy systems \cite{AT1,AT2,Leuzzi,Jamming}.
Finally, the ubiquity of the Monte Carlo method in such far-reaching fields as humanities or finance may precipitate the use of RMI in a host of unforeseen applications,  
such as detecting transitions linked to financial market crashes \cite{market}.

{\em Acknowledgements --} We thank R. Singh, P. Fendley, M. M\"uller, M. Gingras, S. Trebst,  A. Del Maestro, and in particular L. Balents and M. Hastings, for enlightening discussions.  
 This work was made possible by the computing facilities of SHARCNET.  Support was provided 
by NSERC of Canada (A.B.K. and R.G.M.), the Ontario Ministry of Research and Innovation (J.I.), the Vanier Canada Graduate Scholarship program (S.I), and the National Science Foundation under Grant No. NSF PHY11-25915 (R.G.M).

\bibliography{stephen}

\begin{thebibliography}{35}
\expandafter\ifx\csname natexlab\endcsname\relax\def\natexlab#1{#1}\fi
\expandafter\ifx\csname bibnamefont\endcsname\relax
  \def\bibnamefont#1{#1}\fi
\expandafter\ifx\csname bibfnamefont\endcsname\relax
  \def\bibfnamefont#1{#1}\fi
\expandafter\ifx\csname citenamefont\endcsname\relax
  \def\citenamefont#1{#1}\fi
\expandafter\ifx\csname url\endcsname\relax
  \def\url#1{\texttt{#1}}\fi
\expandafter\ifx\csname urlprefix\endcsname\relax\def\urlprefix{URL }\fi
\providecommand{\bibinfo}[2]{#2}
\providecommand{\eprint}[2][]{\url{#2}}

\bibitem[{\citenamefont{Shannon}(1948)}]{Shannon_1948}
\bibinfo{author}{\bibfnamefont{C.~E.} \bibnamefont{Shannon}},
  \bibinfo{journal}{The Bell System Technical Journal}
  \textbf{\bibinfo{volume}{27}}, \bibinfo{pages}{379} (\bibinfo{year}{1948}).

\bibitem[{\citenamefont{Wolf et~al.}(2008)\citenamefont{Wolf, Verstraete,
  Hastings, and Cirac}}]{PhysRevLett.100.070502}
\bibinfo{author}{\bibfnamefont{M.~M.} \bibnamefont{Wolf}},
  \bibinfo{author}{\bibfnamefont{F.}~\bibnamefont{Verstraete}},
  \bibinfo{author}{\bibfnamefont{M.~B.} \bibnamefont{Hastings}},
  \bibnamefont{and} \bibinfo{author}{\bibfnamefont{J.~I.} \bibnamefont{Cirac}},
  \bibinfo{journal}{Phys. Rev. Lett.} \textbf{\bibinfo{volume}{100}},
  \bibinfo{pages}{070502} (\bibinfo{year}{2008}).

\bibitem[{\citenamefont{Wen}(1990)}]{topo_ord}
\bibinfo{author}{\bibfnamefont{X.-G.} \bibnamefont{Wen}},
  \bibinfo{journal}{Int. J. Mod. Phys. B} \textbf{\bibinfo{volume}{4}},
  \bibinfo{pages}{239} (\bibinfo{year}{1990}).

\bibitem[{\citenamefont{Levin and Wen}(2006)}]{LW}
\bibinfo{author}{\bibfnamefont{M.}~\bibnamefont{Levin}} \bibnamefont{and}
  \bibinfo{author}{\bibfnamefont{X.-G.} \bibnamefont{Wen}},
  \bibinfo{journal}{Phys. Rev. Lett} \textbf{\bibinfo{volume}{96}},
  \bibinfo{eid}{110405} (\bibinfo{year}{2006}).

\bibitem[{\citenamefont{Kitaev and Preskill}(2006)}]{KP}
\bibinfo{author}{\bibfnamefont{A.}~\bibnamefont{Kitaev}} \bibnamefont{and}
  \bibinfo{author}{\bibfnamefont{J.}~\bibnamefont{Preskill}},
  \bibinfo{journal}{Phys. Rev. Lett.} \textbf{\bibinfo{volume}{96}},
  \bibinfo{eid}{110404} (\bibinfo{year}{2006}).

\bibitem[{\citenamefont{{Wilms} et~al.}(2012)\citenamefont{{Wilms}, {Vidal},
  {Verstraete}, and {Dusuel}}}]{Wilms1}
\bibinfo{author}{\bibfnamefont{J.}~\bibnamefont{{Wilms}}},
  \bibinfo{author}{\bibfnamefont{J.}~\bibnamefont{{Vidal}}},
  \bibinfo{author}{\bibfnamefont{F.}~\bibnamefont{{Verstraete}}},
  \bibnamefont{and} \bibinfo{author}{\bibfnamefont{S.}~\bibnamefont{{Dusuel}}},
  \bibinfo{journal}{Journal of Statistical Mechanics: Theory and Experiment}
  \textbf{\bibinfo{volume}{1}}, \bibinfo{pages}{23} (\bibinfo{year}{2012}).

\bibitem[{\citenamefont{{Wilms} et~al.}(2011)\citenamefont{{Wilms}, {Troyer},
  and {Verstraete}}}]{Wilms2}
\bibinfo{author}{\bibfnamefont{J.}~\bibnamefont{{Wilms}}},
  \bibinfo{author}{\bibfnamefont{M.}~\bibnamefont{{Troyer}}}, \bibnamefont{and}
  \bibinfo{author}{\bibfnamefont{F.}~\bibnamefont{{Verstraete}}},
  \bibinfo{journal}{Journal of Statistical Mechanics: Theory and Experiment}
  \textbf{\bibinfo{volume}{10}}, \bibinfo{pages}{11} (\bibinfo{year}{2011}).

\bibitem[{\citenamefont{Nakagawa et~al.}(2009)\citenamefont{Nakagawa, Nakamura,
  Motoki, and Zakharov}}]{replicaMC1}
\bibinfo{author}{\bibfnamefont{Y.}~\bibnamefont{Nakagawa}},
  \bibinfo{author}{\bibfnamefont{A.}~\bibnamefont{Nakamura}},
  \bibinfo{author}{\bibfnamefont{S.}~\bibnamefont{Motoki}}, \bibnamefont{and}
  \bibinfo{author}{\bibfnamefont{V.}~\bibnamefont{Zakharov}}
  (\bibinfo{year}{2009}), \eprint{arXiv:0911.2596}.

\bibitem[{\citenamefont{Buividovich and Polikarpov}(2008)}]{Buividovich}
\bibinfo{author}{\bibfnamefont{P.}~\bibnamefont{Buividovich}} \bibnamefont{and}
  \bibinfo{author}{\bibfnamefont{M.}~\bibnamefont{Polikarpov}},
  \bibinfo{journal}{Nuclear Physics B} \textbf{\bibinfo{volume}{802}},
  \bibinfo{pages}{458 } (\bibinfo{year}{2008}).

\bibitem[{\citenamefont{Singh et~al.}(2011)\citenamefont{Singh, Hastings,
  Kallin, and Melko}}]{Roger_MI}
\bibinfo{author}{\bibfnamefont{R.~R.~P.} \bibnamefont{Singh}},
  \bibinfo{author}{\bibfnamefont{M.~B.} \bibnamefont{Hastings}},
  \bibinfo{author}{\bibfnamefont{A.~B.} \bibnamefont{Kallin}},
  \bibnamefont{and} \bibinfo{author}{\bibfnamefont{R.~G.} \bibnamefont{Melko}},
  \bibinfo{journal}{Phys. Rev. Lett.} \textbf{\bibinfo{volume}{106}},
  \bibinfo{pages}{135701} (\bibinfo{year}{2011}).

\bibitem[{\citenamefont{Renyi}(1961)}]{renyi}
\bibinfo{author}{\bibfnamefont{A.}~\bibnamefont{Renyi}},
  \bibinfo{journal}{Proc. of the 4th Berkeley Symposium on Mathematics,
  Statistics and Probability} \textbf{\bibinfo{volume}{1960}},
  \bibinfo{pages}{547} (\bibinfo{year}{1961}).

\bibitem[{\citenamefont{Melko et~al.}(2010)\citenamefont{Melko, Kallin, and
  Hastings}}]{roger_multisheet_sse}
\bibinfo{author}{\bibfnamefont{R.~G.} \bibnamefont{Melko}},
  \bibinfo{author}{\bibfnamefont{A.~B.} \bibnamefont{Kallin}},
  \bibnamefont{and} \bibinfo{author}{\bibfnamefont{M.~B.}
  \bibnamefont{Hastings}}, \bibinfo{journal}{Phys. Rev. B}
  \textbf{\bibinfo{volume}{82}}, \bibinfo{pages}{100409}
  (\bibinfo{year}{2010}).

\bibitem[{\citenamefont{Onsager}(1944)}]{onsager_transfer}
\bibinfo{author}{\bibfnamefont{L.}~\bibnamefont{Onsager}},
  \bibinfo{journal}{Phys. Rev.} \textbf{\bibinfo{volume}{65}},
  \bibinfo{pages}{117} (\bibinfo{year}{1944}).

\bibitem[{\citenamefont{Izmailian et~al.}(2003)\citenamefont{Izmailian,
  Oganesyan, and Hu}}]{Izmailian}
\bibinfo{author}{\bibfnamefont{N.~S.} \bibnamefont{Izmailian}},
  \bibinfo{author}{\bibfnamefont{K.~B.} \bibnamefont{Oganesyan}},
  \bibnamefont{and} \bibinfo{author}{\bibfnamefont{C.-K.} \bibnamefont{Hu}},
  \bibinfo{journal}{Phys. Rev. E} \textbf{\bibinfo{volume}{67}},
  \bibinfo{pages}{066114} (\bibinfo{year}{2003}).

\bibitem[{\citenamefont{Wu et~al.}(2012)\citenamefont{Wu, Izmailian, and
  Guo}}]{Xintian}
\bibinfo{author}{\bibfnamefont{X.}~\bibnamefont{Wu}},
  \bibinfo{author}{\bibfnamefont{N.}~\bibnamefont{Izmailian}},
  \bibnamefont{and} \bibinfo{author}{\bibfnamefont{W.}~\bibnamefont{Guo}},
  \bibinfo{journal}{Phys. Rev. E} \textbf{\bibinfo{volume}{86}},
  \bibinfo{pages}{041149} (\bibinfo{year}{2012}).

\bibitem[{\citenamefont{Inglis and Melko}(2013)}]{stephen_wl}
\bibinfo{author}{\bibfnamefont{S.}~\bibnamefont{Inglis}} \bibnamefont{and}
  \bibinfo{author}{\bibfnamefont{R.~G.} \bibnamefont{Melko}},
  \bibinfo{journal}{Phys. Rev. E} \textbf{\bibinfo{volume}{87}},
  \bibinfo{pages}{013306} (\bibinfo{year}{2013}).

\bibitem[{\citenamefont{McCoy and Wu}(1967)}]{mccoy_wu}
\bibinfo{author}{\bibfnamefont{B.~M.} \bibnamefont{McCoy}} \bibnamefont{and}
  \bibinfo{author}{\bibfnamefont{T.~T.} \bibnamefont{Wu}},
  \bibinfo{journal}{Phys. Rev.} \textbf{\bibinfo{volume}{162}},
  \bibinfo{pages}{436} (\bibinfo{year}{1967}).

\bibitem[{\citenamefont{Au-Yang and Fisher}(1975)}]{me_fisher}
\bibinfo{author}{\bibfnamefont{H.}~\bibnamefont{Au-Yang}} \bibnamefont{and}
  \bibinfo{author}{\bibfnamefont{M.~E.} \bibnamefont{Fisher}},
  \bibinfo{journal}{Phys. Rev. B} \textbf{\bibinfo{volume}{11}},
  \bibinfo{pages}{3469} (\bibinfo{year}{1975}).

\bibitem[{\citenamefont{Kosterlitz and Thouless}(1973)}]{KT_1}
\bibinfo{author}{\bibfnamefont{J.~M.} \bibnamefont{Kosterlitz}}
  \bibnamefont{and} \bibinfo{author}{\bibfnamefont{D.~J.}
  \bibnamefont{Thouless}}, \bibinfo{journal}{Journal of Physics C: Solid State
  Physics} \textbf{\bibinfo{volume}{6}}, \bibinfo{pages}{1181}
  (\bibinfo{year}{1973}).

\bibitem[{\citenamefont{Nelson and Kosterlitz}(1977)}]{NK_jump}
\bibinfo{author}{\bibfnamefont{D.~R.} \bibnamefont{Nelson}} \bibnamefont{and}
  \bibinfo{author}{\bibfnamefont{J.~M.} \bibnamefont{Kosterlitz}},
  \bibinfo{journal}{Phys. Rev. Lett.} \textbf{\bibinfo{volume}{39}},
  \bibinfo{pages}{1201} (\bibinfo{year}{1977}).

\bibitem[{\citenamefont{Hasenbusch}(2005)}]{hasenbusch_TKT}
\bibinfo{author}{\bibfnamefont{M.}~\bibnamefont{Hasenbusch}},
  \bibinfo{journal}{Journal of Physics A: Mathematical and General}
  \textbf{\bibinfo{volume}{38}}, \bibinfo{pages}{5869} (\bibinfo{year}{2005}).

\bibitem[{\citenamefont{Amit et~al.}(1980)\citenamefont{Amit, Goldschmidt, and
  Grinstein}}]{0305-4470-13-2-024}
\bibinfo{author}{\bibfnamefont{D.~J.} \bibnamefont{Amit}},
  \bibinfo{author}{\bibfnamefont{Y.~Y.} \bibnamefont{Goldschmidt}},
  \bibnamefont{and}
  \bibinfo{author}{\bibfnamefont{S.}~\bibnamefont{Grinstein}},
  \bibinfo{journal}{Journal of Physics A: Mathematical and General}
  \textbf{\bibinfo{volume}{13}}, \bibinfo{pages}{585} (\bibinfo{year}{1980}).

\bibitem[{\citenamefont{Shi et~al.}(2011)\citenamefont{Shi, Lamacraft, and
  Fendley}}]{Paul}
\bibinfo{author}{\bibfnamefont{Y.}~\bibnamefont{Shi}},
  \bibinfo{author}{\bibfnamefont{A.}~\bibnamefont{Lamacraft}},
  \bibnamefont{and} \bibinfo{author}{\bibfnamefont{P.}~\bibnamefont{Fendley}},
  \bibinfo{journal}{Phys. Rev. Lett.} \textbf{\bibinfo{volume}{107}},
  \bibinfo{pages}{240601} (\bibinfo{year}{2011}).

\bibitem[{\citenamefont{Nahum et~al.}(2011)\citenamefont{Nahum, Chalker, Serna,
  Ortu\~no, and Somoza}}]{3Dloop}
\bibinfo{author}{\bibfnamefont{A.}~\bibnamefont{Nahum}},
  \bibinfo{author}{\bibfnamefont{J.~T.} \bibnamefont{Chalker}},
  \bibinfo{author}{\bibfnamefont{P.}~\bibnamefont{Serna}},
  \bibinfo{author}{\bibfnamefont{M.}~\bibnamefont{Ortu\~no}}, \bibnamefont{and}
  \bibinfo{author}{\bibfnamefont{A.~M.} \bibnamefont{Somoza}},
  \bibinfo{journal}{Phys. Rev. Lett.} \textbf{\bibinfo{volume}{107}},
  \bibinfo{pages}{110601} (\bibinfo{year}{2011}).

\bibitem[{\citenamefont{Geraedts and Motrunich}(2012)}]{ScottLoop}
\bibinfo{author}{\bibfnamefont{S.~D.} \bibnamefont{Geraedts}} \bibnamefont{and}
  \bibinfo{author}{\bibfnamefont{O.~I.} \bibnamefont{Motrunich}},
  \bibinfo{journal}{Phys. Rev. B} \textbf{\bibinfo{volume}{85}},
  \bibinfo{pages}{144303} (\bibinfo{year}{2012}).

\bibitem[{\citenamefont{Charrier et~al.}(2008)\citenamefont{Charrier, Alet, and
  Pujol}}]{Dimer1}
\bibinfo{author}{\bibfnamefont{D.}~\bibnamefont{Charrier}},
  \bibinfo{author}{\bibfnamefont{F.}~\bibnamefont{Alet}}, \bibnamefont{and}
  \bibinfo{author}{\bibfnamefont{P.}~\bibnamefont{Pujol}},
  \bibinfo{journal}{Phys. Rev. Lett.} \textbf{\bibinfo{volume}{101}},
  \bibinfo{pages}{167205} (\bibinfo{year}{2008}).

\bibitem[{\citenamefont{Chen et~al.}(2009)\citenamefont{Chen, Gukelberger,
  Trebst, Alet, and Balents}}]{Dimer2}
\bibinfo{author}{\bibfnamefont{G.}~\bibnamefont{Chen}},
  \bibinfo{author}{\bibfnamefont{J.}~\bibnamefont{Gukelberger}},
  \bibinfo{author}{\bibfnamefont{S.}~\bibnamefont{Trebst}},
  \bibinfo{author}{\bibfnamefont{F.}~\bibnamefont{Alet}}, \bibnamefont{and}
  \bibinfo{author}{\bibfnamefont{L.}~\bibnamefont{Balents}},
  \bibinfo{journal}{Phys. Rev. B} \textbf{\bibinfo{volume}{80}},
  \bibinfo{pages}{045112} (\bibinfo{year}{2009}).

\bibitem[{\citenamefont{Swingle and Senthil}(2011)}]{Brian}
\bibinfo{author}{\bibfnamefont{B.}~\bibnamefont{Swingle}} \bibnamefont{and}
  \bibinfo{author}{\bibfnamefont{T.}~\bibnamefont{Senthil}}
  (\bibinfo{year}{2011}), \eprint{arXiv:1109.3185}.

\bibitem[{\citenamefont{Castelnovo et~al.}(2010)\citenamefont{Castelnovo,
  Chamon, and Sherrington}}]{ClaudioGlass}
\bibinfo{author}{\bibfnamefont{C.}~\bibnamefont{Castelnovo}},
  \bibinfo{author}{\bibfnamefont{C.}~\bibnamefont{Chamon}}, \bibnamefont{and}
  \bibinfo{author}{\bibfnamefont{D.}~\bibnamefont{Sherrington}},
  \bibinfo{journal}{Phys. Rev. B} \textbf{\bibinfo{volume}{81}},
  \bibinfo{pages}{184303} (\bibinfo{year}{2010}).

\bibitem[{\citenamefont{Young and Katzgraber}(2004)}]{AT1}
\bibinfo{author}{\bibfnamefont{A.~P.} \bibnamefont{Young}} \bibnamefont{and}
  \bibinfo{author}{\bibfnamefont{H.~G.} \bibnamefont{Katzgraber}},
  \bibinfo{journal}{Phys. Rev. Lett.} \textbf{\bibinfo{volume}{93}},
  \bibinfo{pages}{207203} (\bibinfo{year}{2004}).

\bibitem[{\citenamefont{Katzgraber et~al.}(2009)\citenamefont{Katzgraber,
  Larson, and Young}}]{AT2}
\bibinfo{author}{\bibfnamefont{H.~G.} \bibnamefont{Katzgraber}},
  \bibinfo{author}{\bibfnamefont{D.}~\bibnamefont{Larson}}, \bibnamefont{and}
  \bibinfo{author}{\bibfnamefont{A.~P.} \bibnamefont{Young}},
  \bibinfo{journal}{Phys. Rev. Lett.} \textbf{\bibinfo{volume}{102}},
  \bibinfo{pages}{177205} (\bibinfo{year}{2009}).

\bibitem[{\citenamefont{Leuzzi et~al.}(2009)\citenamefont{Leuzzi, Parisi,
  Ricci-Tersenghi, and Ruiz-Lorenzo}}]{Leuzzi}
\bibinfo{author}{\bibfnamefont{L.}~\bibnamefont{Leuzzi}},
  \bibinfo{author}{\bibfnamefont{G.}~\bibnamefont{Parisi}},
  \bibinfo{author}{\bibfnamefont{F.}~\bibnamefont{Ricci-Tersenghi}},
  \bibnamefont{and} \bibinfo{author}{\bibfnamefont{J.~J.}
  \bibnamefont{Ruiz-Lorenzo}}, \bibinfo{journal}{Phys. Rev. Lett.}
  \textbf{\bibinfo{volume}{103}}, \bibinfo{pages}{267201}
  (\bibinfo{year}{2009}).

\bibitem[{\citenamefont{Coniglio and Nicodemi}(2000)}]{Jamming}
\bibinfo{author}{\bibfnamefont{A.}~\bibnamefont{Coniglio}} \bibnamefont{and}
  \bibinfo{author}{\bibfnamefont{M.}~\bibnamefont{Nicodemi}},
  \bibinfo{journal}{J. Phys.: Cond. Matt.} \textbf{\bibinfo{volume}{12}},
  \bibinfo{pages}{6601} (\bibinfo{year}{2000}).

\bibitem[{\citenamefont{Harre and Bossomaier}(2009)}]{market}
\bibinfo{author}{\bibfnamefont{M.}~\bibnamefont{Harre}} \bibnamefont{and}
  \bibinfo{author}{\bibfnamefont{T.}~\bibnamefont{Bossomaier}},
  \bibinfo{journal}{Europhysics Letters} \textbf{\bibinfo{volume}{87}},
  \bibinfo{pages}{18009} (\bibinfo{year}{2009}).

\bibitem[{\citenamefont{{Humeniuk} and {Roscilde}}(2012)}]{roscilde_qmc}
\bibinfo{author}{\bibfnamefont{S.}~\bibnamefont{{Humeniuk}}} \bibnamefont{and}
  \bibinfo{author}{\bibfnamefont{T.}~\bibnamefont{{Roscilde}}},
  \bibinfo{journal}{ArXiv e-prints}  (\bibinfo{year}{2012}),
  \eprint{1203.5752}.

\end{thebibliography}

\end{document}